\documentclass[printer]{aa} % for a referee version
\usepackage{graphicx}
\usepackage{epstopdf}
\usepackage{txfonts}
\usepackage{multirow}
\usepackage{longtable}
\begin{document}
   \title{A southern hemisphere survey of the 5780 and 6284 \AA\ diffuse interstellar bands: correlation with the extinction\thanks{Based on observations collected at the European Southern
Observatory, La Silla, Chile.}}

   %\subtitle{}

   \author{S. Raimond \inst{1}
   	\and
          R. Lallement \inst{2}
           \and
          J.L. Vergely \inst{3}
          \and
          C. Babusiaux\inst{2}
          \and
          L. Eyer\inst{4}
     }

   \institute{  1 - Universit\'e Versailles Saint-Quentin, LATMOS/IPSL, 11 Bd d'Alembert, 78200 Guyancourt, France\\
               2 - GEPI/ Paris Observatory, 5 Place Jules Janssen, 92195 Meudon, France\email{rosine.lallement@obspm.fr} \\
            3-  ACRI-ST, 260 route du Pin Montard, Sophia-Antipolis, France\\
            4- Observatoire de Gen\`eve, Universit\'e de Gen\`eve, Chemin des Maillettes 51, CH-1290 Sauverny, Switzerland
}

   \date{Received ; revised }

% \abstract{}{}{}{}{}
% 5 {} token are mandatory

  \abstract
  % context heading (optional)
  % {} leave it empty if necessary
   {}
  % aims heading (mandatory)
   {Diffuse interstellar bands (DIBs) measured in stellar spectra contain information on the amount of interstellar (IS) matter that is distributed along the line-of-sight, and similarly to other absorbing species may be used to locate IS clouds. Here we present a new database of 5780.5 and 6283.8 \AA\ DIB measurements. Those two DIBs have the advantage that they are strong and also broad enough to be detectable in cool-star spectra. We also study their correlation with the reddening.}
  % methods heading (mandatory)
   {The database is based on high-resolution, high-quality spectra of  early-type nearby stars located in the southern hemisphere at an average distance of 300 pc.  Equivalent widths of the two DIBs were determined by means of a realistic continuum fitting and synthetic atmospheric transmissions. For all stars that possess a precise measurement of their color excess, we compare the DIBs and the extinction.}
  % results heading (mandatory)
   {We find average linear relationships of the DIBS and the color excess based on $\simeq$ 120 and 130 objects that agree well with those of a previous survey of $\simeq$ 130 northern hemisphere stars closer than 550pc. Because our target sky coverage is complementary, this similarity shows that there is no significant spatial dependence of the average relationship in the solar neighborhood within $\simeq$600 pc.  A noticeably different result is our higher degree of correlation of the two DIBs with the extinction, especially for the 5780\AA\ DIB. We demonstrate that it is simply due to the lower temperature and intrinsic luminosity of our targets. Using cooler target stars reduces the number of {\it outliers}, especially for nearby stars, confirming that the radiation field of UV bright stars has a significant influence on the DIB strength. We illustrate the potential use of 3D maps of the ISM for characterizing the DIB sites. There is some evidence that interstellar cavity boundaries are DIB-deficient, although definite conclusions  will have to wait for maps with a higher resolution. Finally, we have used the cleanest data to compute updated DIB shapes.\thanks{available from the CDS, Strasbourg}}
  % conclusions heading (optional), leave it empty if necessary
   {}

   \keywords{ISM; diffuse interstellar bands; interstellar dust}

   \maketitle
%
%________________________________________________________________

\section{Introduction}

Identifying the diffuse interstellar bands (DIBs), the $\simeq$ 400 weak absorption features seen in the spectra of reddened stars (\cite{hobbs2008}; 2009)    remains one of the longest-standing spectroscopic problems in astrophysics, and today they are still widely studied with the aim of conclusively identifying their carriers in the interstellar medium (ISM). If large molecules are preferentially designated (\cite{legdhend85}), nothing has been decisively established as yet on their actual structures and sizes, on their participation from the gaseous or solid phase, on the role of the charge state balance, etc.. (for summaries see \cite{herbig95}; \cite{JD94}; \cite{salama96}; \cite{fulara00}; \cite{snow11}; \cite{fried11} (hereafter FR11) and references therein). Clues to their origin have been investigated in different ways. Laboratory spectra and models of rotational excitation of molecules have been compared to observations with the aim of finding convincing matches (see the recent review by \cite{sarre2006}). Polarization studies aim at constraining the carrier phase, with most results pointing against the solid state (\cite{smith77}; and recently \cite{cox2006}; 2011).

At the same time, the DIBs are studied toward specific targets or for large statistical samples of stars with the aim of establishing correlations with other interstellar tracers and identifying {\it families} among the DIBS that are characterized by similar correlations. Sightline categories have been identified, in particular the so-called $\sigma$- and $\zeta$-type sightlines which have specific associations of DIB properties and UV extinction laws (\cite{kre94}). $\zeta$-type sightlines  correspond to UV-shielded cloud cores while $\sigma$-type sightlines probe cloud external regions that are partially ionized by the UV radiation field. The influence of the radiation field has been recently studied by Vos et al. (2011) based on stars from the upper Scorpius subgroup of the Sco OB2 association. These authors established significant differences between the properties of the two groups in terms of DIB-DIB, DIB-extinction, and DIB-gas relationships.

These relationships have also been found based on sky-distributed datasets: links between the strengths of some DIBs and HI,  H$_{2}$, KI, NaI or C$_{2}$ columns, between some of the DIBs and the dust column traced by the color excess, or between DIBS themselves  have been found  by several authors (e.g. \cite{krelo87}; \cite{thor03}; \cite{welty06}; \cite{ fried11}; \cite{mccall2010}). The influence of the UV radiation field of the target stars has again found to be significant. In addition to
the potential destruction of the carriers by energetic photons, the combination of the charge state and the carrier size is thought to play a major  role in a number of DIBs (\cite{gala04}; \cite{salama96}). 
%Thorburn et al (2003) have identified a series of narrow DIBs that are correlated with C$_{2}$

% article salama et al: conditions champ detremine etats de charge, or les gros et petits ne donnent des trnasix et dib que sous certaines condic de charges: donc ca change de falimlles via la taille, mais en raison de l'ionisation

% ajouter krelowski ehrenfreund cox-boulanger

Instead of focusing on DIB properties and their differential behaviors that may give clues to their carriers, it is a more practical aspect that has motivated the present statistical work. Although correlations between the DIB strengths and the color excesses or the gas columns may be weak,  DIBS are carrying some information on the quantity of matter along the path to a target star. In the same way that neutral sodium has been used to build local ISM maps (\cite{welsh10}; \cite{vergely10}), despite the absence of a strong correlation between neutral sodium and H columns, DIB absorption strengths toward stars at increasing distances can be used to determine IS cloud locations. Forthcoming extensive stellar datasets and hopefully parallaxes from the ESA Gaia mission will open the way to constructing precise 3D maps of the galactic dust by means of the inversion method (\cite{vergely1}). These maps will in turn be useful for the interpretation of the stellar surveys, by helping to break the degeneracy  between the star temperature and the reddening. Our aim is to build new derivation methods of the DIBS in all stellar spectra, to help constraining the 3D mapping, and in parallel to study the relationships between the DIB strengths and other IS parameters in more detail and to cover all various ISM cloud types and locations in the Galaxy.

Here we complement previous statistical studies of two strong DIBS by an analysis of  southern hemisphere high-quality, high-spectral resolution data, and in particular to compare the results with the recent results of an extensive northern hemisphere survey presented by FR11, and of a survey dedicated to the Upper Scorpius area (\cite{Vos11}). While the FR11 survey aimed at detailed studies of DIB-to-DIB pairwise correlations, as part of a search for DIB carrier identification, our data  were recorded with the aim of building a three-dimensional map of the nearby ISM. This is why the choice of the targets is drastically different (see section 4). Because we are using nearby targets with low reddening, we focused on two of the strongest bands in the red domain. Both have been  studied by several authors, and in particular  by  FR11. They are those that probably are the most readily extractable from survey data such as the Gaia-ESO spectra, especially for cool stars, because the DIBS in this case must be broader than the stellar lines.

In section 2 we describe the spectroscopic dataset used throughout the paper. In section 3, we explain the cleaning and fitting techniques that led to the extraction of the DIBs. We also describe a new method to estimate the error induced by fitting a continuum to both sides of the DIB.
In section 4 we describe the correlative studies based on this new dataset, and the combination with the previous measurements of FR11. In section 5 we  discuss our results and some potential improvements, in particular, attempts to identify the ISM environments that lead to substantial deviations from the average trends.

\section{Observations}

Our measurements were extracted from a database of 500 high signal-to-noise (S/N $\geq$100), high-resolution (R$\simeq$48,000) spectra of nearby stars, mostly located within 400 parsecs, acquired with the FEROS spectrograph at the ESO/Max Planck 2.2m telescope in La Silla  (LP179.C-0197 program, total 15 nights, ended February 2009). The aim of this survey was developing an interstellar gas absorption database and subsequently producing of a three-dimensional map of the local ISM. Interstellar neutral sodium and singly ionized calcium have been measured in those spectra by means of profile-fitting techniques and combined with previous existing data to provide local maps (\cite{vergely10}; \cite{ welsh10}). To facilitate interstellar line identification and fitting, the spectral types range from early B to A5, with fast rotators preferentially selected for the latest types (see the distribution of stellar types in Fig. 1). 

These high-quality data are also perfectly suited for spectroscopic studies of the strongest diffuse bands, provided the target stars are sufficiently reddened for the DIBS to be detectable. Fig. \ref{examples} shows typical examples of the spectral intervals containing the $\lambda\lambda$ 5780.4 and 6283.8 diffuse bands. Thanks to the weakness of the stellar lines and the resulting smooth continua, the 5780.4\AA\ DIB can be extracted very easily when it dominates the noise, with the exception of some of the latest stellar type targets for which three stellar lines contaminate the DIB spectral interval. On the other hand, the 6284.8\AA\ DIB is very strongly contaminated by telluric molecular oxygen  lines, but thanks to the high resolution and the sharpness of those lines the DIB is measurable, provided one correctly removes the telluric absorptions. 

This dataset is also ideally adapted to statistical studies. The target stars are randomly distributed in the sky and in distance because of mapping requirements. Fig. 1 shows the location of the target stars that possess precise extinction data (see below). Superimposed are the Friedman et al. (2011)  targets that formed the basis of their correlative studies. Evidently, the two stellar datasets are spatially complementary, which is essentially due to the locations of the two observatories, the northern hemisphere Apache Point and the southern hemisphere La Silla Observatories. This complementarity is used here to test the variability of the DIBS with location.
Our survey is restricted to targets with measurable DIBs and a reddening determination (see section 5). Because the difficulties associated with the two DIBS are of a different origin (stellar lines vs telluric lines, see next section), for a few stars only one of the two DIBS was measured.

\begin{figure}[h]
	\begin{center}
		\includegraphics[width=8cm,height=6cm,angle=-90]{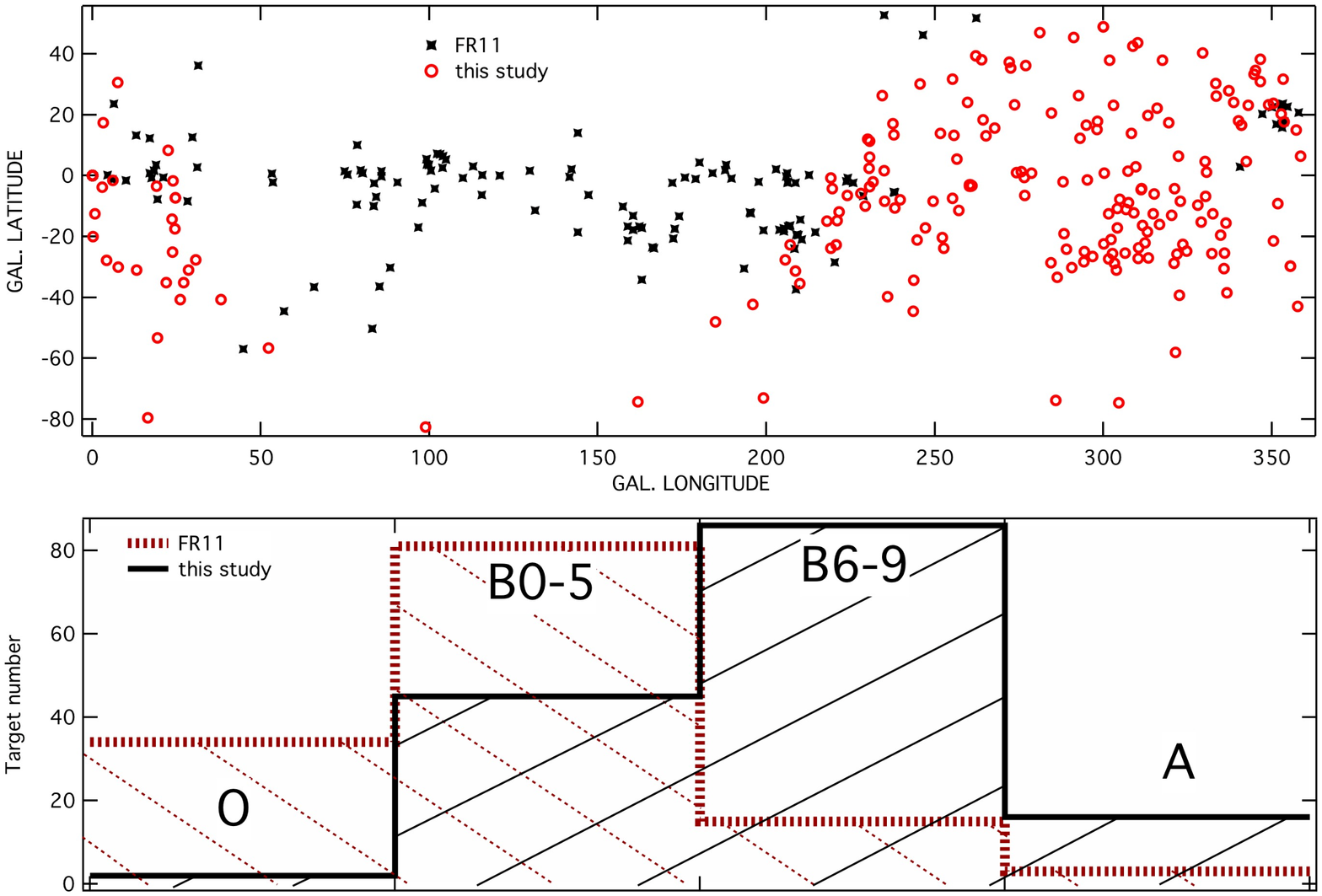}
\caption{Location of our FEROS targets (red circles), and comparison with the FR11 survey targets (black crosses). Also shown are the distributions of stellar types (bottom).}
\end{center}
\label{sky_type}
\end{figure}

\begin{table*}
\caption{Stellar data and measurements. Color excesses are from Geneva photometry. The full table is available as online material.}
\begin{center}
\begin{tabular}{|l|c|c|c|c|c|c|c|c|c|}
\hline
Star & l ($^{\circ}$) & b ($^{\circ}$) & spectral type & d (pc) & E(B-V) * & EW(5780A) (m\AA\ )  & EW(6284A) (m\AA\ ) \\
\hline
HD385 & 303.81 & -31.01 & B9IV & 287 & 0.065 & $82.7 \pm 3.0$ & $197.6 \pm 4.6$ \\
\hline
HD955 & 78.89 & -77.09 & B4V & 316 & 0.006 & & $35.3 \pm 7.5$ \\
\hline
HD1348 & 303.25 & -28.74 & B9.5IV & 549 & 0.106 & $101.6 \pm 2.8$ & $302.9 \pm 5.3$ \\
\hline
HD4751 & 304.57 & -74.56 & B8V& 292 & -0.007 & $12.5 \pm 2.0$ & $13.1 \pm 2.2$ \\
\hline
HD7795 & 285.89 & -73.74 & B9III/IV & 232 & 0.025 & $11.5 \pm 2.0$ & \\
\hline
HD9399 & 161.91 & -74.19 & A1V& 231 & 0.044 & $3.9 \pm 3.0$ & $7.4 \pm 3.4$ \\
\hline
HD12561 & 199.10 & -73.04 & B6V& 204 & 0.009 & $28.3 \pm 2.3$ & \\
\hline
HD20404 & 184.93 & -47.95 & B8 & 296 & 0.017 & $33.6 \pm 2.0$ & $75.3 \pm 4.0$ \\
\hline
HD24446 & 196.02 & -42.31 & B9 & 602 & 0.024 & $14.0 \pm 2.4$ & $36.2 \pm 4.3$ \\
\hline
HD30397 & 235.96 & -39.75 & A0V& 203 & 0.002 & $15.0 \pm 3.0$ & \\
\hline
HD30963 & 208.57 & -31.24 & B9 & 240 & 0.016 & $20.4 \pm 1.5$ & \\
\hline
HD32043 & 205.56 & -27.57 & B9 & 338 & 0.057 & $45.9 \pm 1.9$ & $45.4 \pm 4.0$ \\
\hline
HD33244 & 286.28 & -33.38 & B9.5V& 273 & 0.050 & $39.8\pm 3.0$ & $85.0 \pm 4.1$ \\
\hline
HD37104 & 219.11 & -23.74 & B5IV/V & 292 & 0.026 & $18.2 \pm 1.1$ & \\
\hline
HD37971 & 220.74 & -22.75 & B4/B5III & 565 & 0.030 & $6.2 \pm 1.5$ & \\
\hline
HD38602 & 290.68 & -30.22 & B8III& 262 & 0.100 & $52.1\pm 3.0$ & $185.8 \pm 4.8$ \\
\hline
HD41814 & 218.04 & -14.90 & B3V& 338 & 0.021 & $15.3 \pm 1.7$ & \\
\hline
HD42849 & 220.94 & -14.70 & B9.5III& 351 & 0.049 & $37.2 \pm 2.4$ & $54.3 \pm 4.0$ \\
\hline
HD44533 & 284.47 & -28.52 & B8V& 292 & 0.064 & $22.5 \pm 3.0$ & $99.8 \pm 5.5$ \\
\hline
HD44737 & 252.62 & -23.84 & B7V& 746 & 0.010 & $24.3 \pm 1.8$ & \\
\hline
HD44996 & 221.58 & -11.80 & B4V& 289 & 0.095 & $53.5 \pm 1.8$ & $130.3 \pm 3.8$ \\
\hline
HD45040 & 294.17 & -28.30 & B9IV/V & 196 & 0.066 & $31.3\pm 3.0$ & $96.5 \pm 6.6$ \\
\hline
HD45098 & 244.76 & -21.11 & B5V& 565 & 0.039 & $31.9 \pm 1.8$ & \\
\hline
HD46976 & 279.25 & -27.05 & B9V& 364 & 0.014 & & $5.3 \pm 4.5$ \\
\hline
HD48150 & 252.27 & -20.25 & B3V& 485 & 0.046 & $12.9 \pm 1.8$ & \\
\hline
HD48872 & 229.32 & -9.99 & B5III/IV & 336 & 0.044 & $22.9 \pm 1.7$ & $60.3 \pm 3.5$ \\
\hline
HD49336 & 247.13 & -17.18 & B4Vne& 407 & 0.050 & $10.1 \pm 2.2$ & \\
\hline
HD49481 & 219.50 & -4.21 & B8 & 365 & 0.033 & $36.0 \pm 1.9$ & \\
\hline
HD49573 & 224.07 & -6.47 & B8II/III & 395 & 0.060 & $67.4 \pm 2.1$ & $172.3 \pm 3.0$ \\
\hline
HD51876 & 228.05 & -5.75 & B9IIw& 333 & 0.064 & $30.0 \pm 1.6$ & $141.7 \pm 4.0$ \\
\hline
HD52266 & 219.13 & -0.68 & O9V& 552 & 0.295 & $179.1 \pm 3.1$ & $518.4 \pm 5.0$ \\
\hline
HD52849 & 235.09 & -8.32 & B3IV & 1351 & 0.029 & $15.5 \pm 2.2$ & \\
\hline
HD55523 & 239.64 & -7.95 & B3III& 373 & 0.020 & $18.7 \pm 2.3$ & \\
\hline
HD57139 & 231.65 & -1.97 & B5II/III & 357 & 0.149 & $99.4 \pm 2.8$ & $179.7 \pm 4.4$ \\
\hline
HD60098 & 249.46 & -8.29 & B4V& 248 & 0.043 & $42.5 \pm 2.4$ & \\
\hline
HD60102 & 296.77 & -26.51 & B9.2/A0V & 207 & 0.085 & $34.1 \pm 3.0$ & $119.2 \pm 6.1$ \\
\hline
HD60325 & 230.45 & 2.52 & B2II & 617 & 0.180 & $112.8 \pm 3.3$ & $254.5 \pm 9.7$ \\
\hline
HD60929 & 257.11 & -11.45 & A0V& 196 & 0.014 & $22.5 \pm 2.3$ & \\
\hline
HD61554 & 234.92 & 1.61 & B6V& 257 & 0.072 & $47.4 \pm 2.0$ & $114.1 \pm 4.3$ \\
\hline
HD63112 & 230.59 & 6.17 & B9III& 228 & 0.034 & $35.2 \pm 1.8$ & $106.9 \pm 2.8$ \\
\hline
HD63868 & 255.17 & -7.41 & B3V& 327 & 0.033 & $19.6 \pm 2.3$ & \\
\hline
HD65322 & 301.53 & -27.26 & B8IV & 224 & 0.085 & $53.9\pm 3.0$ & $199.9 \pm 6.5$ \\
\hline
HD70948 & 260.67 & -3.45 & B5V& 341 & 0.052 & $21.7\pm 3.0$ & $55.9 \pm 6.5$ \\
\hline
HD71019 & 260.37 & -3.14 & B3II/III & & 0.095 & $75.1\pm 3.0$ & $200.9 \pm 5.4$\\
\hline
HD71123 & 260.22 & -2.92 & B9III& 415 & 0.062 & $44.0\pm 3.0$ & $47.0 \pm 5.5$ \\
\hline
HD71336 & 261.01 & -3.21 & B3III/IV & & 0.055 & $23.8\pm 3.0$ &  \\
 \hline
\end{tabular}
\end{center}
\begin{footnotesize}
(*) The error on E(B-V) is assumed to be 0.03 mag
\end{footnotesize}
\label{default}
\end{table*}

%\addtocounter{table}{1}
%\begin{table}[htdp]
%\begin{table}[htdp]
%\begin{table*}

\begin{figure}[!h]
\begin{minipage}[t]{0.3\linewidth}
\centering
  	\includegraphics[width=1\linewidth]{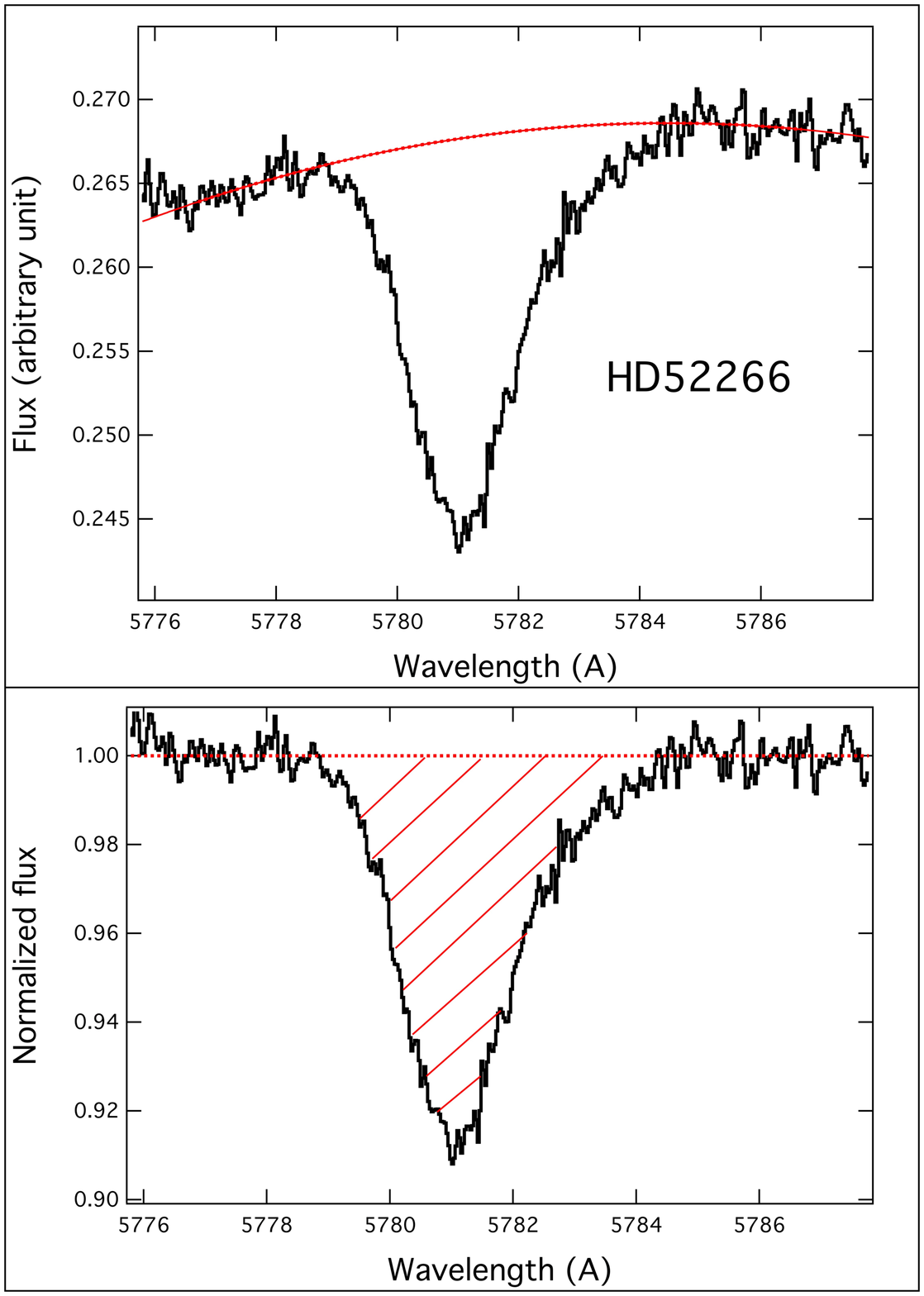}
\end{minipage}\hfill
\begin{minipage}[t]{0.3\linewidth}
\centering
  	\includegraphics[width=1\linewidth]{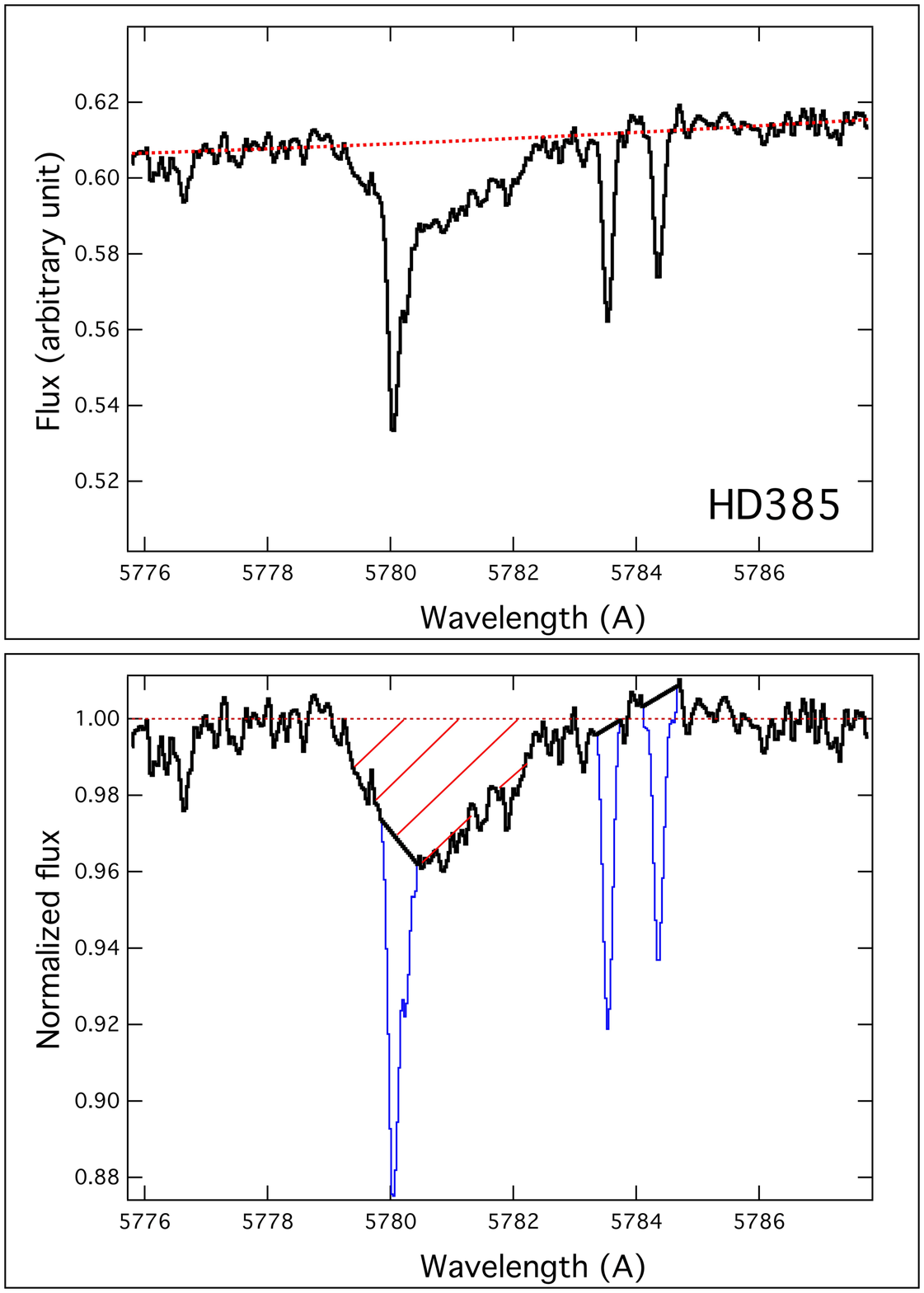}
\end{minipage}\hfill
\begin{minipage}[t]{0.3\linewidth}
\centering
  	\includegraphics[width=1\linewidth]{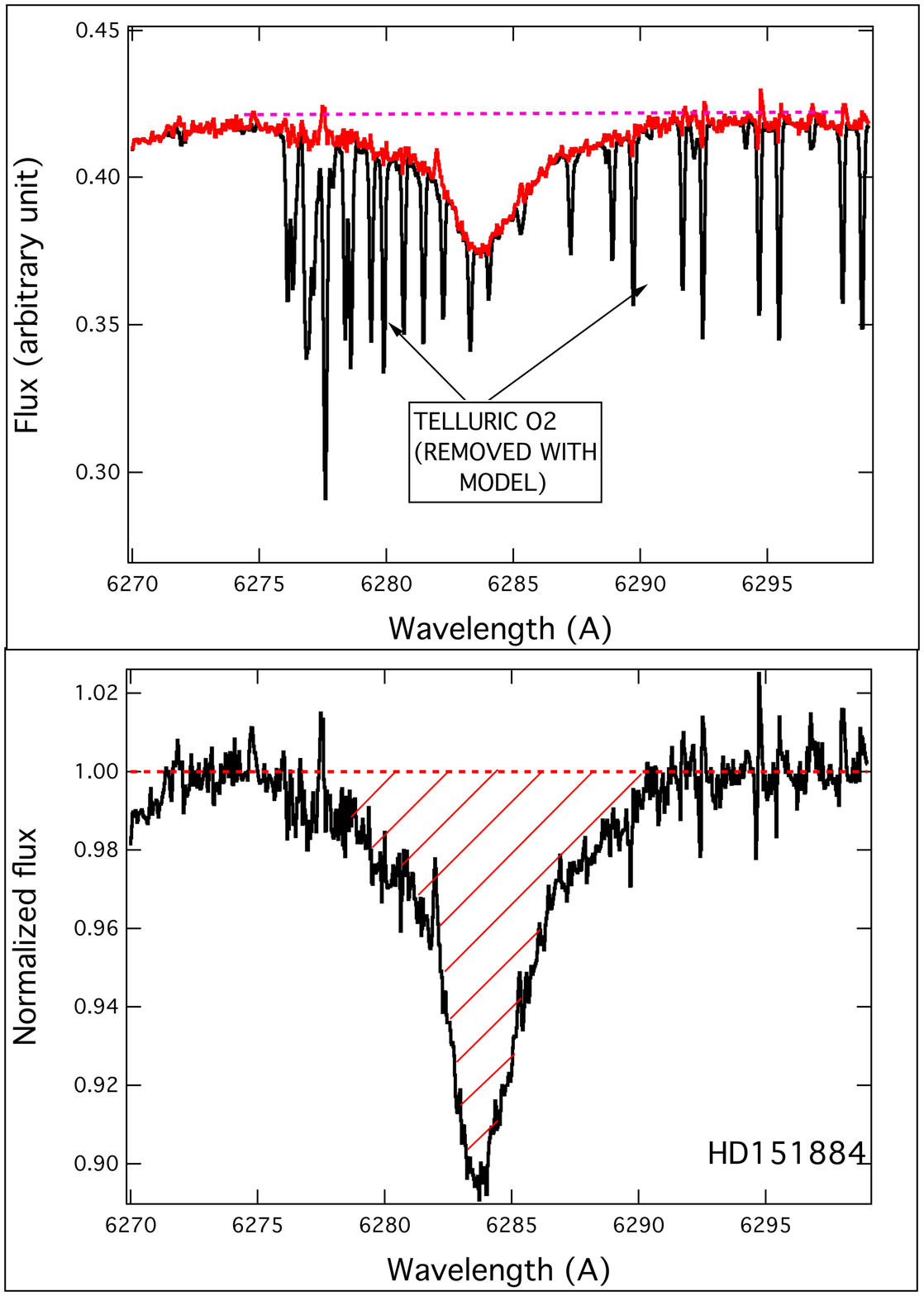}
\end{minipage}
\caption{Illustrations of the DIB equivalent width measurements. The top (bottom) graphs show the spectra before (after) normalization, i.e., division by the fitted continuum. The hatched areas correspond to the measured equivalent widths. (Left) A typical continuum fitting and EW extraction of the 5780\AA\ DIB. (Middle) Example of removing contaminating stellar lines for the late-B target stars. (Right) Example of telluric line removal by means of a synthetic transmission for the 6284\AA\ DIB and the subsequent EW measurement.}
\label{examples}
\end{figure}

\section{Data analysis}

\subsection{The 5780\AA\ diffuse band}

The 5780\AA\ absorption is one of the strongest DIBs and has been already measured toward bright stars at various distances. It is about 3.5\AA\ wide (full width) and, as said above, is contained in a spectral region that is mostly free of stellar lines for the majority of our targets (Fig 2), with the exception of three FeII stellar lines that appear in late-B objects and may be strong enough to contaminate the band significantly. The 5780\AA\  DIB has been shown by FR11 to be fairly well correlated with the neutral gas column and less well correlated with the extinction. Vos et al. (2011) show that for targets of  the Upper Scorpius OB association the DIB is significantly weaker in $\zeta$-type sightlines (i.e. UV-shielded central parts of the IS clouds).

The equivalent width (EW) is classically measured by fitting the stellar continuum on both sides of the main DIB absorption (see Fig. \ref{examples}). Following FR11, we considered the 5778.0 to 5784\AA\  interval.  The EW was computed based on the normalized spectrum. For all targets we carefully searched for the potential contaminating stellar lines, using the fact that the third line at 5784.45 \AA\  is shifted out of the DIB and can be easily detected (see Fig. \ref{examples}). When the lines are too broad and strong to permit an accurate measurement of the DIB, we excluded  the corresponding target. When they are narrow enough to allow a reliable correction, as in the example shown in Fig. \ref{examples}, we applied a mask as illustrated in the figure and replaced the spectrum by a straight line within the spectral intervals of the stellar features. The error on the DIB equivalent width associated with this procedure is conservatively estimated  to be less than 10\% of the removed stellar line EW for the whole dataset.

The target stars and the derived equivalent widths are listed in Table 1. Among the 500 targets, about 200 stars have measurable $\lambda\lambda$ 5780.4 diffuse bands. For a signal-to-noise (S/N) of $\simeq100$, our minimum equivalent width and error is about 3 m\AA\  and is essentially caused by uncertainties on the shape of the stellar continuum in this area. This limit is reflected in the data point dispersion at very low reddening values.  
For the targets that are too close and whose ISM column is too small, the corresponding threshold is about E(B-V) $\simeq$ 0.05 (see Fig. \ref{dibcorr1}). 

\subsection{The 6284\AA\ diffuse band}

According to FR11, the broad 6284\AA\ band is slightly less correlated with HI and also the reddening compared to the 5780\AA\ DIB. 
We used a synthetic telluric transmission to correct for the narrow lines that contaminate the 6284\AA\ DIB spectral region (see Fig. \ref{examples}). This transmission was computed using the LBLRTM code (Line-By-Line Radiative Transfer Model, \cite{lblrtm05}) and the HITRAN (HIgh-Resolution TRANsmission molecular absorption, \cite{hitran2008}) spectroscopic database for a standard atmospheric profile and the altitude of La Silla.  For each star we modified the atmospheric transmission to take into account the terrestrial line Doppler shift and the instrumental width and adjusted it to the airmass by means of an automatic fitting routine. In this way, all lines vary in the same proportions, i.e., we did not allow a differential variability among the various lines. Because a small differential variability is present for some lines, the removal of the entire series of telluric lines is not perfect (see Fig. \ref{examples}), but it is by far sufficient for our purpose. The adopted correction is the one for which the length (defined as the sum of distances between consecutive points) of the residual spectrum obtained after division of the initial spectrum by the transmission, reaches a minimum.

The 6284\AA\ DIB is on average twice as strong as than the 5780\AA\ DIB. However, it is also significantly broader, and finally  shallower on average than the 5780\AA\ absorption. Moreover, the telluric line removal procedure introduces some noise that also tends to reduce its detectability. This explains why, among all targets, only about 130 stars have measurable $\lambda\lambda$ 6284 diffuse bands. For a S/N of $\simeq100$, our limit for the equivalent width is about 3 m\AA\, similar to the 5780\AA\ DIB. The derived equivalent widths are listed in Table 1. The interval for the DIB that is considered here is the same as in FR11, namely 6275-6292\AA\ .

\section{Error estimates}

Uncertainties on the equivalent width measurements have two main origins. A first source of uncertainty is linked to the signal-to-noise and the width of the DIB and can be calculated in a straightforward way. For the 5780 and 6284\AA\ DIBs those errors are about 3 and 5 to10 m\AA\ , respectively. The error on the second DIB is larger as a consequence of the broader width and the telluric line correction. The second source of uncertainty is the continuum adjustment and is much more difficult to estimate. It depends on the large-scale features in the spectrum that may have various origins: imperfections of the reduction, fringes, stellar features, presence of lines from a companion, etc. Anything that renders the spectrum different from a very smooth curve that can be extrapolated from the spectral intervals on both sides of the DIB may lead to a slight error in the continuum shape (esp. the curvature) and a subsequent error on the EW. 

We devised and performed a new method to estimate these uncertainties that we call the {\it sliding windows error estimate}, based on the assumption that the departures from the smooth continuum adjusted to the DIB sides that are present at the DIB location must also be present elsewhere in the spectrum. Within this assumption, one  way to estimate potential errors on the continuum placement is to use a large part of the spectrum and perform continuum fittings similar to those used at the DIB location, but this time at many locations without DIB absorptions. By similar, we mean based on the same wavelength intervals that are used on the two sides of the DIB. This change of locations is made by simply shifting the intervals all along the spectrum. For each of the {\it sliding windows} locations a continuum fitting is performed, and the equivalent width comprised between the fitted continuum and the actual spectrum (which should be null if the fitted continuum were perfectly adjusted) is computed and stored. The entire set of  locations provides many error values, from which we extracted a standard deviation. 

We combined  the two errors quadratically and the results are quoted in Table 1. Because the error obtained when using the {\it sliding windows} method is in most cases larger than the first error caused by the noise only, the final error is close the {\it sliding windows}  value.  We note that while all continua were adjusted {\it by eye} with different polynomial laws and spectral intervals from one star to the other, the quoted errors were been obtained by means of a unique automatic code applied to all stars, for a linear continuum and two 3\AA\ wide windows. Since this method provides a very conservative value, we believe it is representative of maximum errors associated to the actual determinations. For a few stars that have a cold companion, the automated method provides unrealistically high values because it integrates sharp stellar lines. In this case we  used the maximum error obtained for all other objects.

\section{Correlations with the color excess E(B-V)}

\begin{figure}[!h]
\centering
\includegraphics[width=1\linewidth,height=9cm]{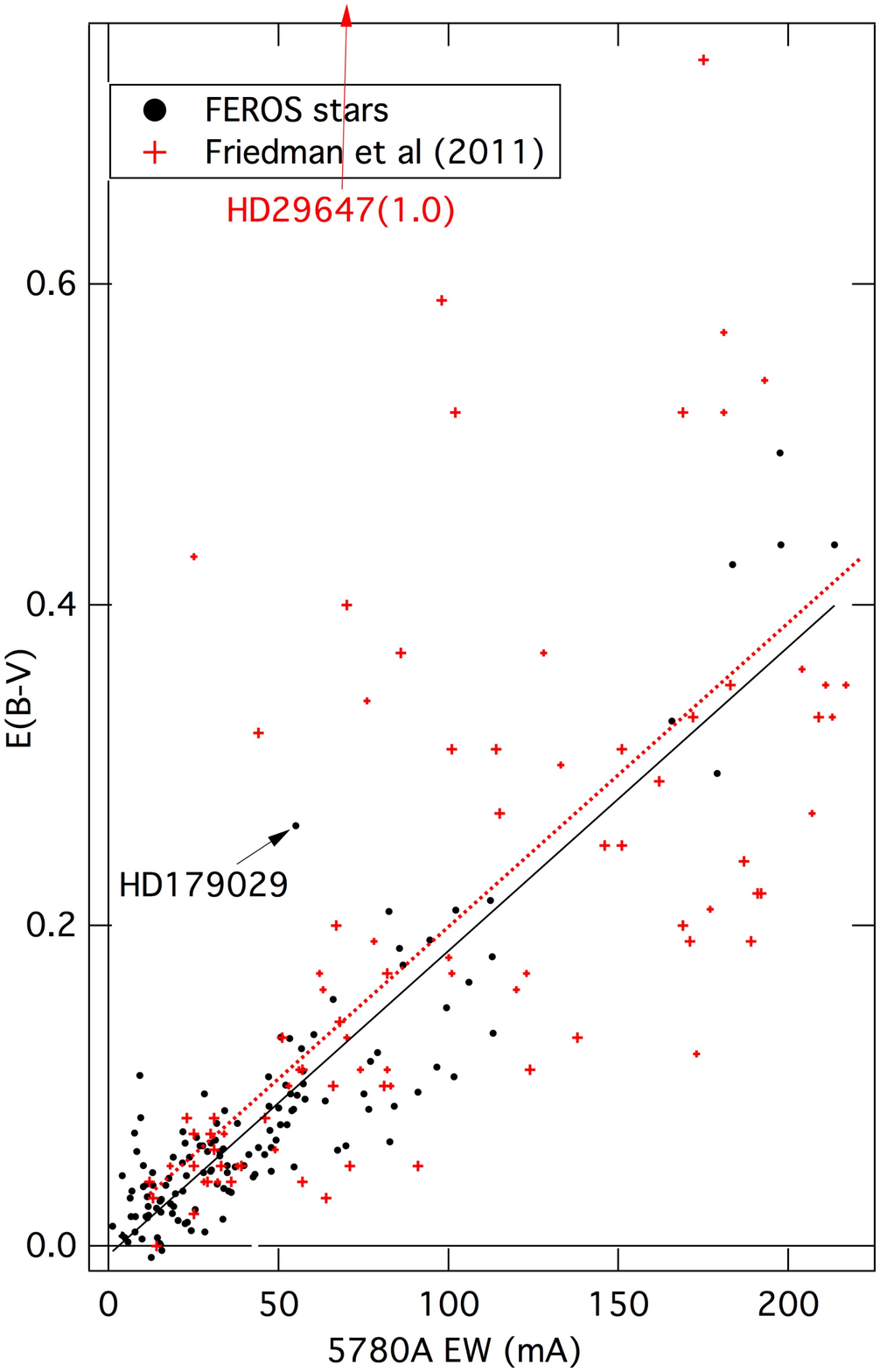}
\caption{Johnson color excess as a function of the 5780\AA\ DIB equivalent width for our FEROS stars (black filled circles). Also shown are the results of FR11 (red plus signs) for equivalent widths lower than 230 m\AA\  . Stars beyond 400 parsecs are indicated by small markers. Dashed and solid lines are the linear correlations for the FR11 et FEROS data. The FEROS correlation corresponds to the ODR method and all data. The FR11 correlation is for case (3) of Table 2. }
\label{dibcorr1}
\end{figure}

\begin{figure}[!h]
\centering
\includegraphics[width=1\linewidth,height=9cm]{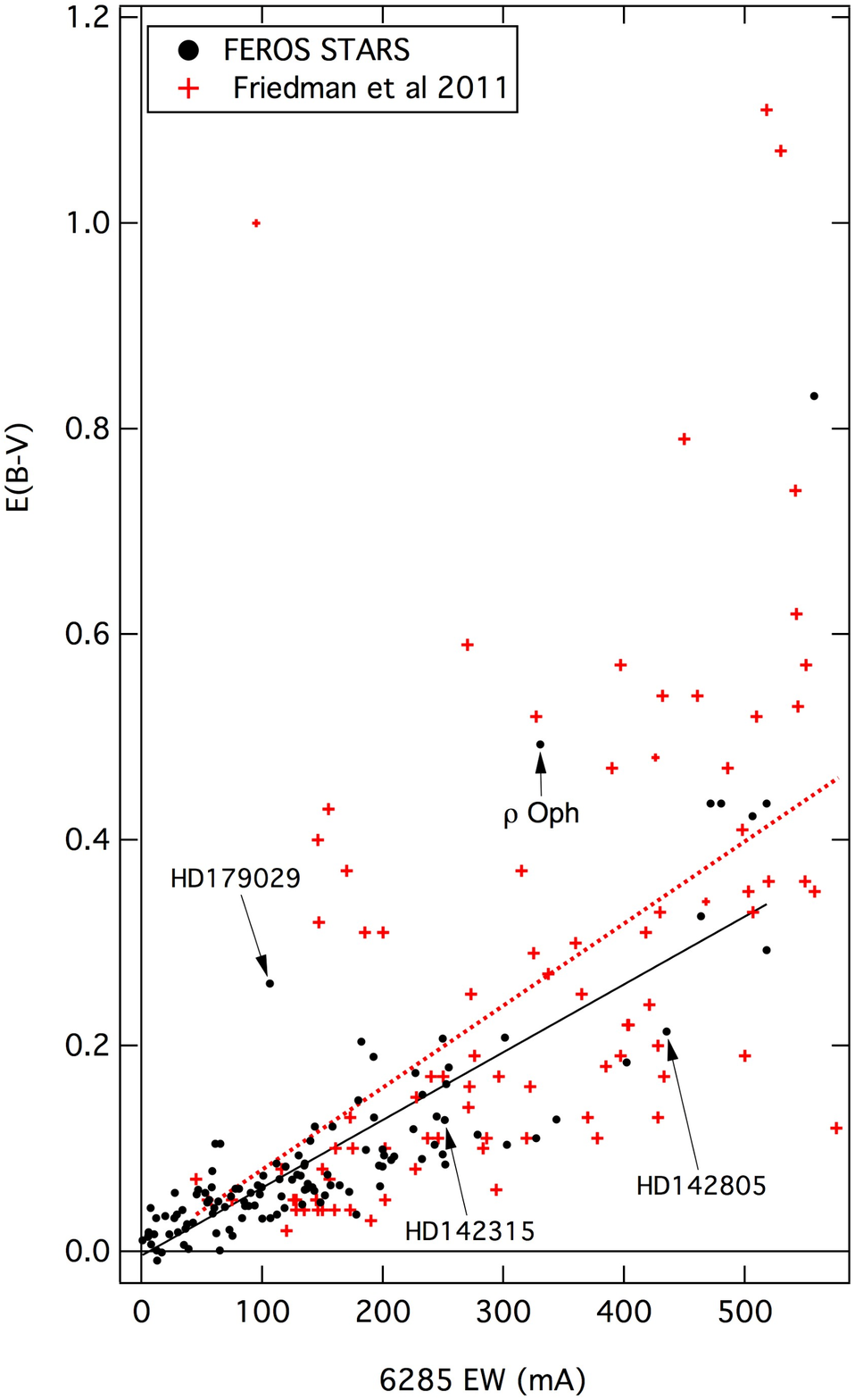}
\caption{Same as Fig. \ref{dibcorr1} for the 6284 DIB.}
\label{dibcorr2}
\end{figure}

For the sake of homogeneity, we restricted the correlation to a set of targets for which the extinction was determined based on the same photometric system and calibration method. We used here the extensive database of bright stars observed in the seven-colors Geneva photometric system and the associated color excess determinations (\cite{cram99}; \cite{burcram12}). One hundred and thirty-five  targets simultaneously possess an E(B-V) determination and a measurable 5780\AA\ DIB, while 120 targets possess an E(B-V) and a measurable 6283.8\AA\ DIB. Because the Geneva photometric system is slightly different from the other systems, we cross-correlated the Geneva color excess values with the Str\"{o}mgren E(b-y) determinations for all common targets  and derived a linear correspondency between those quantities. The relationship between the Johnson E(B-V) and the Str\"{o}mgren E(b-y) was taken from \cite{vergely10}. In Table 1 are listed the color excess values, along with the target star galactic coordinates and the DIB equivalent width measurements. We kept the error on the color excess that is associated to the calibration itself, which excludes any dispersion due to stellar intrinsic variabilities that cannot be taken into account.  Those errors are fairly small, while the dispersion as well as the existence of negative values at low reddening indicate actual larger uncertainties. We thus conservatively estimated an error on E(B-V) of 0.03 in the linear regression discussed below. Figures \ref{dibcorr1} and \ref{dibcorr2} display the color excess as a function of the EW for the two DIBS. 

We assumed that the DIB absorptions are sufficiently weak for us to be still in the linear regime, and that the DIB EW is on average proportional to the extinction. In establishing the correlations we used two methods, (i) a linear fit that does not take into account the measured individual error bars, but estimates errors on the parameters based on the observed data point dispersion, and (ii) a linear fit with the
Orthogonal Distance Regression (ODR) method (also called “total
least-squares method”) that takes into account the individual estimated errors on both the extinction values and the DIB EWs and minimizes the weighed orthogonal distance from the data to the fitted
curve. We used the ODRPACK95 package (\cite{boggs89})
 implemented in the Wavemetrics/IGOR 6.0 software. Two noticeable facts can be derived from the figure: first, the data point dispersion for low values of E(B-V)s and EWs is higher than the EW uncertainties would suggest. We believe that the uncertainty on the color excess is partly, but only partly, responsible for this dispersion (see our discussion below). Second, there are conspicuous {\it outliers} in this plot, particularly a few data points that have a weak diffuse band while the color excess and the average fitted linear law imply a stronger one. It is known that such very low DIB strengths may exist (e.g. \cite{porce92}).
For both DIBS the fit coefficients from the two methods and the Pearson correlation coefficient (wich is independent of the error bars) are presented in Table 2. As can be seen in the table, the ODR-fitted parameters are very similar to the parameters that emerge in the absence of error bar weighting. For clarity, Figs. \ref{dibcorr1} and \ref{dibcorr2} show the ODR fit only. 
Also shown are the correlation parameters and coefficient obtained when excluding targets for which the departure from the mean relationship for E(B-V) is larger than 0.1 (rep 0.15) mag, which corresponds to one (resp. two) outliers for the 5780 (res. 6284) DIB. One of the outliers is $\rho$Oph, also excluded from the correlation by FR11. The second outlier is HD179029, a B5V star that has peculiarly low DIB values (see the discussion below).

We compared our results with the results of FR11 for these two bands, using their table 1. The FR11 data points are superimposed on our FEROS determinations, and linear fits are similarly shown. Note that their data points, which correspond to the strongest EWs, i.e. largely above the FEROS  measurements, are not visible in Figs. \ref{dibcorr1} and \ref{dibcorr2}. Table 2 lists the fit parameters and Pearson correlation coefficient taken from Table 4 of FR11.

Our linear relationship coefficients are very close to those found by FR11 for the 5780\AA\ DIB, i.e., these two independent databases generally lead to the same dependence of the DIB  with reddening. Because the two surveys do not correspond to the same hemisphere,
this implies that there is no obvious dependence on the galactic direction, at least for the nearby ($\leq$ 500 pc) ISM . There is a slightly larger
 difference between the FR11 coefficients and our numbers for the 6284\AA\ DIB. Data points with a low DIB and a high E(B-V) are more numerous in the FR11 dataset, which may influence the intercept value. While it is negligible in our case, it is found to be E(B-V) (at DIB=0) = -0.02 for FR11.
Comparing with the Vos et al. (2011) results on nearby ISM in Upper Scorpius, our average slope of 525 mA/E(B-V) for the 5780\AA\  DIB is slightly above their average slope of  460 mA/E(B-V). 

For the 5780\AA\ DIB, the FEROS Pearson correlation coefficient of 0.92 appears significantly higher (see table 2) than the coefficient found by FR11 (0.82). This difference becomes even larger if we limit  the FR11 data to the same range of EWs as in our FEROS data, (i.e. EW $\leq$ 220 m\AA\ ),  which reduces the Pearson coefficient to 0.62 only. We have examined the potential reasons for this difference. First, it is not caused by the use of different sources for the color excess. We arrived at this conclusion after comparing of the E(B-V) quoted by FR11 and the corresponding Geneva determination for the 60 FR11 target stars that are included in the Geneva catalog. The dispersion around the linearity is one order of magnitude lower (at least) than the E(B-V) vs EW dispersion. Second, we considered a potential effect of target location distribution in the sky. Fig. \ref{sky_type}  shows that there are no obvious differences in the way the targets are distributed, with both datasets possessing low- and high-latitude stars, and no conspicuous concentrations in specific regions, which makes this influence unlikely. This agrees with the similarity between the two mean relationships. 

\begin{figure}[!h]
\centering
  	\includegraphics[width=1\linewidth,height=8cm]{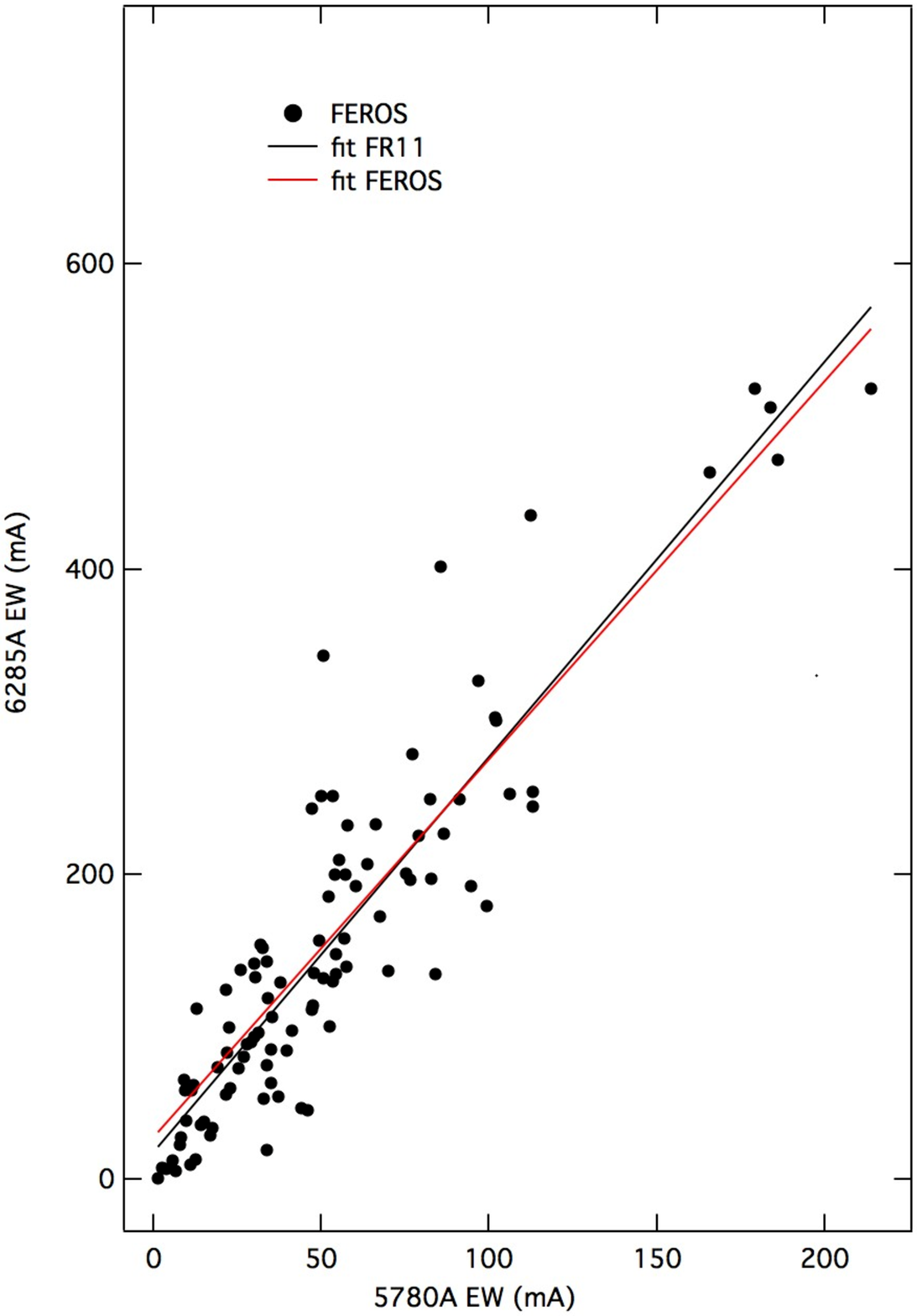}

\caption{Internal correlation between the two DIBs. Also shown is the DIB-DIB correlation found by FR11.}
\label{dibdib}
\end{figure}

Our smaller dispersion can certainly not be caused by a better data quality or a more precise EW computation. The contrary is demonstrated by the comparison between the internal correlation coefficients derived from the two datasets. The correlation between the FEROS 5780 and 6284 DIBS equivalent widths is shown in Fig. \ref{dibdib}. Our 5780 vs 6284 Pearson correlation coefficient is 0.91, i.e., inferior to the 0.96 correlation coefficient derived by FR11, which implies the absence of an additional dispersion due to their EW derivation. We believe that our higher dispersion for the DIB-DIB correlation is due to the superior resolution and signal-to-noise of the FR11 data, and to the fact that our EWs are smaller.  Indeed, the 6284-5780 coefficient of FR11 decreases from 0.96 down to 0.86 when restricting the DIB-DIB correlation to 5780\AA\ EWs smaller than 22m\AA\ . We also note some small differences on the internal correlations. Our average linear relationship is found to be

EW (6284) = (18.6  $\pm$ 8.17) + (2.59 $\pm$ 0.12) * EW (5780),

which denotes a bias toward a lowest 6284 threshold. These numbers are different from the results of FR11:  a= 28.2 $\pm$ 5.8 and  b= 2.32 $\pm$ 0.03. Imposing the same intercept of 28.2, we derive b=2.48 $\pm$ 0.08, which is already much closer to their relationship, and suggests that there are indeed biases toward positive intercepts linked to some outliers with low 5780\AA\ DIBS.  
 
We then investigated the effect of the target distances. Our targets are nearby stars, 80\% are located within 400 pc and only 3 stars lie between 650 and 1150 pc (Hipparcos parallaxes). The FR11 targets are on average more distant, with only 50\% of targets within 400 pc and $\simeq$ 50 stars beyond 650 pc. In principle, the dispersion should decrease with distance, because cloud-to-cloud variations are averaged. However, to improve the comparison between our datasets, we fitted a linear relationship between the FR11 reddening values and their 5780.4 EWs for a restricted set of targets closer than 400 pc. In the same way that the coefficient decreases with the EW threshold, it also decreases by introducing such a threshold on the distance  (0.62 Pearson coefficient  instead of 0.82). This implies that the reason for a tighter correlation in our dataset is not linked to the choice of closer targets. 

We finally considered the stellar characteristics in both samples: here there is a marked difference that is a direct consequence of the difference between the goals of the two surveys. Fig. 1 shows the distribution of stellar types. While FR11 mostly used very bright and hot stars (mostly earlier than B5) to obtain spectra of extremely high quality, the present sample is mainly composed of moderately hot stars, with a maximum around B7, chosen for their location in direction and distance, and does not contain bright blue stars because such stars, being easy targets, have already been observed in the past with other instruments. Moreover, about half of the FR11 targets are giants or subgiants (class I to III), while our targets are mainly dwarfs ($\simeq$ 20\% of class I-III). Given this difference, our interpretation is the following: owing to the effect of the UV radiation field on the two DIBs, and according to recent findings (e.g. \cite{Vos11}), there is a strong variability of their concentration or of their charge state around the hot blue stars that are part of the FR11 study. For nearby stars, the fraction of the matter along the path length that is influenced by the hot target star may be large, e.g. if the densest cloud along the sightline is close to it (or is engulfing the target). For this reason, the total DIB absorption strongly reflects this environment, hence the variability of the DIB to dust ratio. For a set of moderately hot target stars, this effect is not so much pronounced and the line-of-sight to line-of-sight variability is not as high. This explains our tighter correlation and the correlation strength increase with distance for the FR11 dataset.  It is evidently linked also to the larger number of "outliers"  in the FR 11 data that are discussed and thought to be linked to high-radiation environments.

\section{Perspectives}

Our study and the performed comparisons shed additional light on the correlation between the 5780\AA\ and 6284\AA\ DIBs and the reddening, as well as on the influence of the stellar radiation field on those DIBs. The fact that average relationships are similar for the two hemispheres shows that those relationships do not vary much with location. The diminution of the dispersion when using late-type targets, especially for the  5780\AA\ DIB, is useful in the sense that it helps in approaching the most appropriate average coefficients to use for conversion purposes, if necessary. Our best-fit scaling for the 5780\AA\ DIB corresponds to 525 mA/E(B-V). It is very close to an average of the two 419 and 640 mA/E(B-V) coefficients found by \cite{Vos11} for the Upper Scorpius $\sigma$-type (or external ionized regions) and $\zeta$-type (shielded cloud cores) sightlines, respectively. 

\begin{figure}[!h]
\includegraphics[width=1\linewidth,height=5 cm,angle=-90]{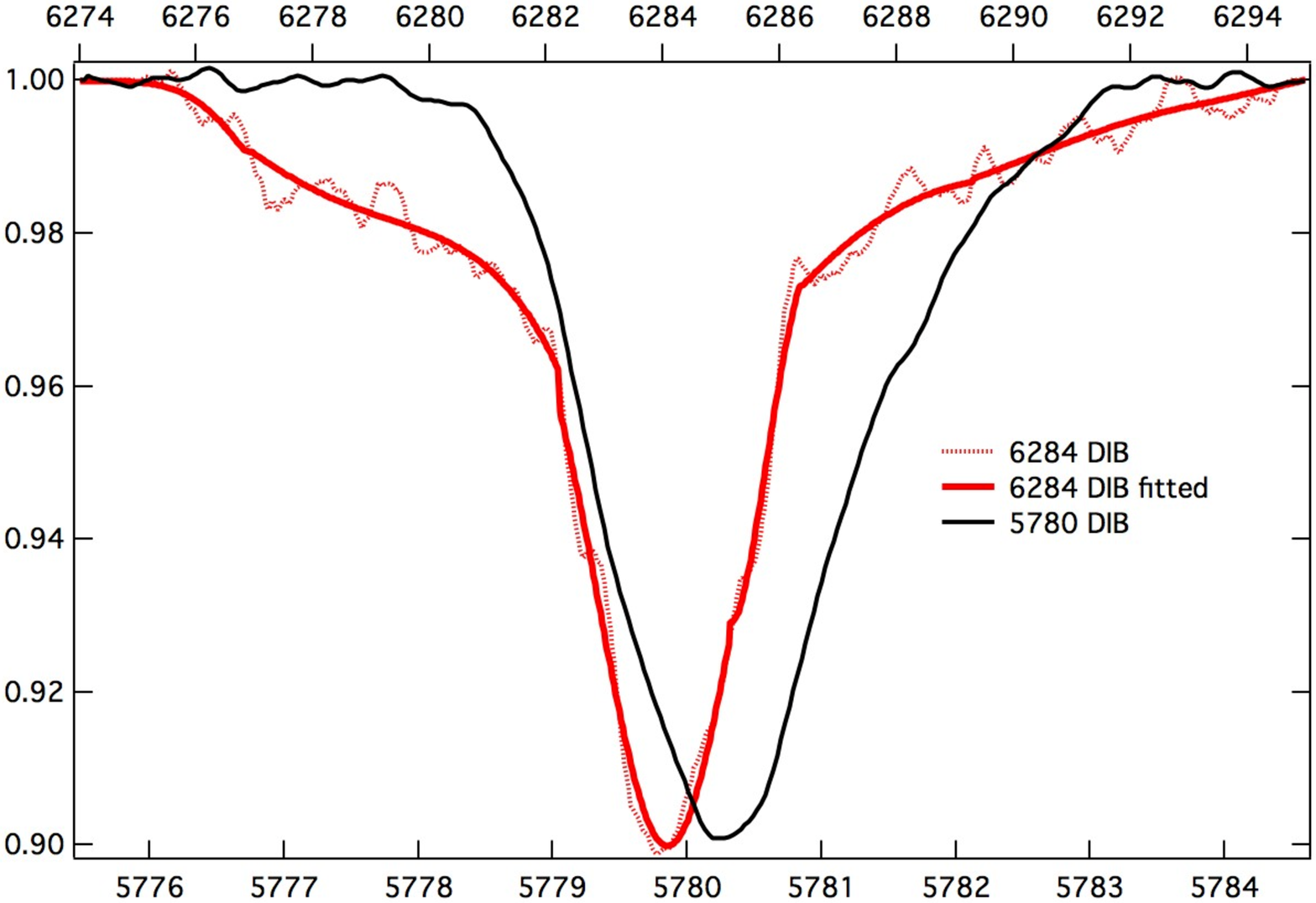}
\caption{Average DIB shapes derived from our FEROS spectra. For the 6284 DIB, we show a smoothed profile where the small-scale structures that remain after the averaging process were removed (see text).  The three curves are available on-line from the CDS.}
\label{temp_dibs}
\end{figure}

We used the then strongest and cleanest measured DIBs, normalized them, scaled them to the same equivalent width and averaged them to provide their most precise shapes for use at resolutions lower than 50,000. The results are shown in Fig. \ref{temp_dibs}. For the 6284\AA\ DIB there are irregularities that remain despite the averaging process. We believe they are caused by imperfections of the telluric line removal procedure, but it is difficult to preclude some substructures of the DIB itself. We provide both the raw shape and a smoothed version where those substructures have been removed. The 5780\AA\ DIB corresponds to an EW of 217 m\AA\ ( E(B-V) $\simeq$  0.41 with the average relationship) while the 6284\AA\ DIB corresponds to EW=500 m\AA\, and E(B-V) $\simeq$ 0.32.

More studies at larger distance will allow one to investigate the evolution of the DIB-extinction relationship, and better constrain the DIB sites. 
In this respect, the building of 3D maps of galactic dust or gas based on extinction and absorption data will allow progress in characterizing the DIB production sites. Figs. 5 and 6 show cuts in a 3D opacity distribution inverted from extinction measurements in the solar neighborhood (\cite {tarantola}; \cite{vergely10}) that contain sightlines with average or peculiar DIB to extinction ratios. The one to the {\it outlier} HD179029 is tangential to a nearby cavity
presumably  blown by stellar winds or supernovae, and does not cross any large dense cloud association. The two {\it normal} sightlines toward HD142805 and HD142315 both cross an extended high-density area, devoid of very bright early-type stars. 
These maps have a low spatial resolution of about 15 pc, which is a
strongly limiting factor for any study of the line-of-sight characterization. In future, studying with more detailed maps the relationships between the DIB to reddening ratios as a function of such sightline characteristics will hopefully provide additional clues to the DIB formation sites.

\begin{table*}[!htdp]
\caption{Correlations with E(B-V)}
\begin{center}
\begin{tabular}{|l|c|c|c|c|c|c|c|c|c|}
\hline
DIB & a(*) & b(*) & mA/E(B-V) (**) & Pearson correl. coeff. & target number & dataset charact. \\
5780\AA\ & -0.006$\pm$ 0.004 & 1.91$\pm$0.08 10$^{-3}$ &  &0.91 & 135 & FEROS(all)  \\
"& -0.011$\pm$ 0.005 & 2.04$\pm$0.08 10$^{-3}$&  &  & " & FEROS(all) ODR \\
"& -0.007$\pm$ 0.004 & 1.90$\pm$0.07 10$^{-3}$ & 525 $\pm$ 25 & 0.92 & 134 & FEROS (1) \\
"& -0.011$\pm$ 0.005 & 2.01$\pm$0.08 10$^{-3}$ & 498 $\pm$ 25 & & " & FEROS (1) ODR \\
"&  0.01 & 1.9810$^{-3}$  & 505 $\pm$ 3 &  0.82 & 129 & FR11 (3) \\
"&  0.05 & 1.510$^{-3}$  & &  0.61 & 85 & FR11 (£) \\
"&   &  & 419&   & & Vos et al (2011)  $\zeta$-type\\
"&  &   & 640&  &  &   "                 $\sigma$-type\\
\hline
6284\AA\ & -0.004$\pm$ 0.007 & 6.59$\pm$0.39 10$^{-4}$ &  & 0.84 & 120 & FEROS (all)  \\
" &  -0.011 $\pm$0.005 & 6.98$\pm$ 0.25 10$^{-4}$  &&" & " & FEROS (all) ODR \\
" &  -0.004 $\pm$ 0.006 & 6.33$\pm$0.32 10$^{-4}$ & 1588 $\pm$ 80  & 0.88 & 118 & FEROS (2) \\
" &  -0.008 $\pm$0.005 & 6.49$\pm$ 0.25 10$^{-4}$  & 1539 $\pm$ 50 & " & " & FEROS (2) ODR \\
"  & -0.018 & 8.210$^{-4}$ && 0.81 & 125 & FR11 (3) \\
" & -0.022 & 8.710$^{-4}$ && 0.62 & 89 & FR11 (4) \\
\hline

 \end{tabular}
\end{center}
\begin{footnotesize}
(*) for a linear fit E(B-V) = a+ b*EW . a is in magnitude units (mag), b is in mag/m\AA\
(**) Average equivalent width per reddening unit (m\AA\ / mag)
(1) excluding HD179029
(2) excluding $\rho$Oph and HD179029
(3) excluding $\rho$Oph, HD29647, HD37061, $\theta$1OriC
(£) excluding $\rho$Oph, HD29647, HD37061, $\theta$1OriC,  and restricted to stars with EW$\leq$ 230 m\AA\ 
(4) excluding $\rho$Oph, HD29647, HD37061, $\theta$1OriC,  and restricted to stars with EW$\leq$ 600 m\AA\ 
\end{footnotesize}
\label{default}
\end{table*}%

\begin{figure}[!h]
\begin{center}
\includegraphics[width=\linewidth,angle=-90]{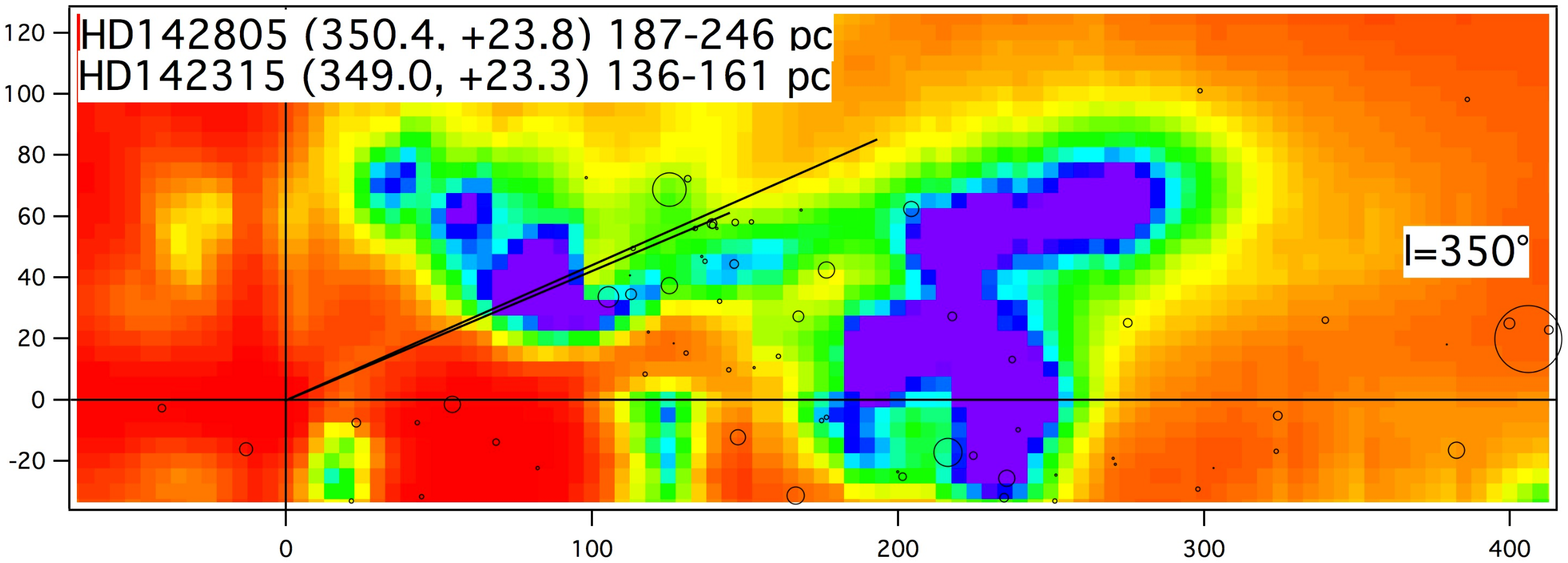}
\caption{Cut in a 3D opacity cube derived by inversion of color excess data (Vergely et al. 2010), in a vertical plane containing the Sun and the two stars HD142805 and 142315. Distances are in parsecs. The Sun is at (0,0) coordinates. The North Galactic Pole corresponds to the Y axis and the X axis is in the Plane toward the longitude 350 degrees. Violet represents the densest clouds, while red corresponds to the lowest density. A black line joining the Sun and the star allows one to appreciate the ISM distribution along the line-of-sight.  Most of the intervening interstellar matter (violet) corresponds to dense clouds and there are no bright hot stars close to the LOS. This may be linked to their  {\it normal} 5780\AA\ DIB strength. The black circles show the UV stars from the Hipparcos catalog and the circle diameter scales with their ionizing power}
\end{center}
\label{maps_1}
\end{figure}

\begin{figure}[!h]
\begin{center}
		\includegraphics[width=\linewidth,height=5.5cm]{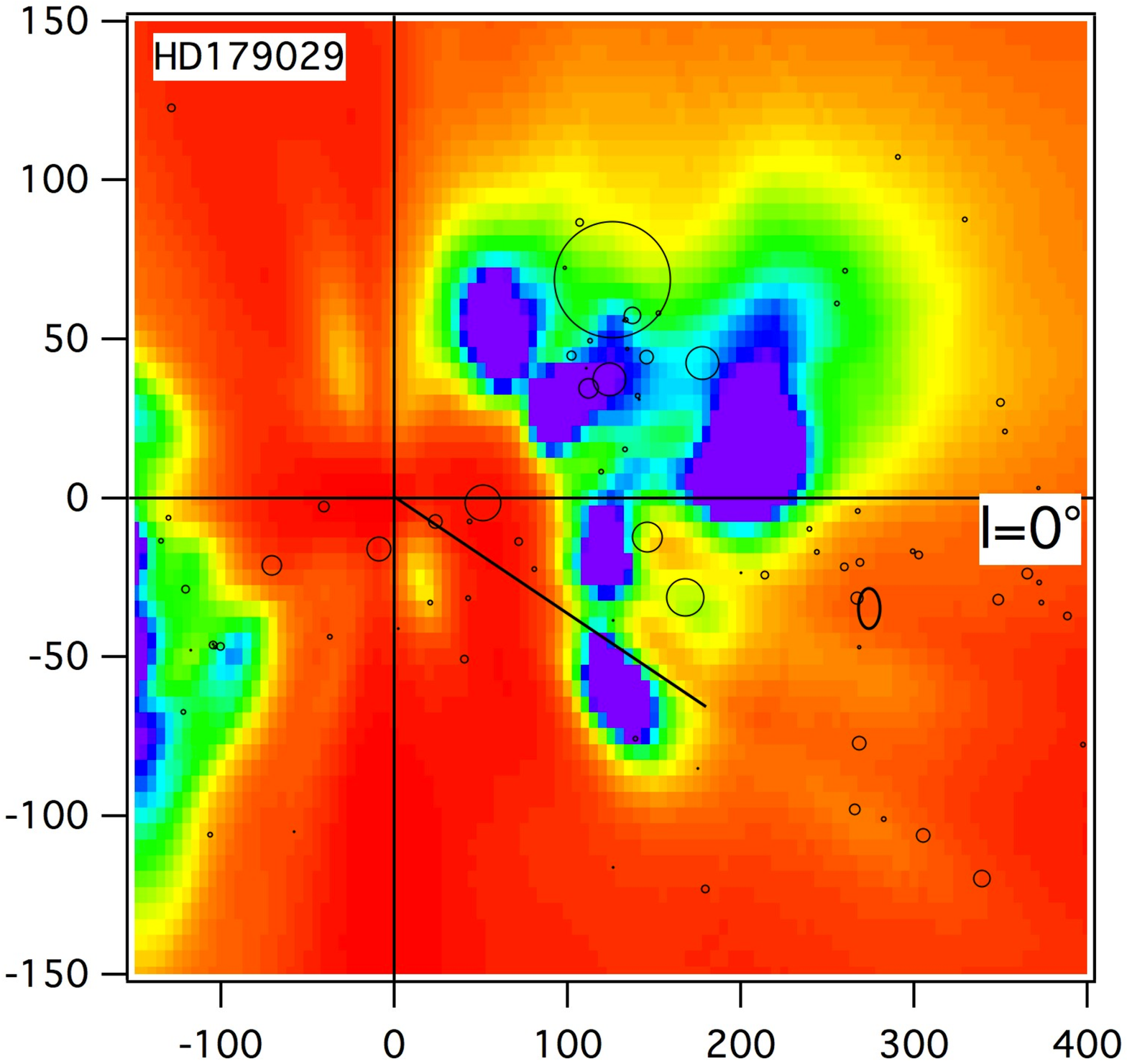}
\caption{Same as figure 5 for the outlier HD179029. The path to the star corresponds to the boundary of a dense complex. A large part of the ISM along this LOS corresponds to the periphery of tenuous, presumably hot gas cavities and is not shielded from the ambient radiation field (the Local Cavity and a second cavity beyond 150 pc). }
\end{center}
\label{maps_2}
\end{figure}

\begin{acknowledgements}

We deeply thank Lucky Puspitarini, who coded the {\it sliding windows} method for equivalent width error estimates and applied it to all spectra.
\end{acknowledgements}

\Online
%\begin{appendix}

\section{Online Material: Stellar data and DIB measurements}

\onltab{1}{
\begin{table*}
\caption{Stellar data and measurements. Color excesses are from Geneva photometry. }
\begin{center}
\begin{tabular}{|l|c|c|c|c|c|c|c|c|c|}
\hline
Star & l ($^{\circ}$) & b ($^{\circ}$) & spectral type & d (pc) & E(B-V) * & EW(5780A) (m\AA\ )  & EW(6284A) (m\AA\ ) \\
\hline
HD385 & 303.81 & -31.01 & B9IV & 287 & 0.065 & $82.7 \pm 17.0 $ & $197.6 \pm 11.0$ \\
\hline
HD955 & 78.89 & -77.09 & B4V & 316 & 0.006 & & $35.3 \pm 24.8 $ \\
\hline
HD1348 & 303.25 & -28.74 & B9.5IV & 549 & 0.106 & $101.6 \pm 13.2 $ & $302.9 \pm 21.2 $ \\
\hline
HD4751 & 304.57 & -74.56 & B8V& 292 & -0.007 & $12.5 \pm 8.4 $ & $13.1 \pm 22.4 $ \\
\hline
HD7795 & 285.89 & -73.74 & B9III/IV & 232 & 0.025 & $11.5 \pm 8.0 $ & \\
\hline
HD9399 & 161.91 & -74.19 & A1V& 231 & 0.044 & $3.9 \pm 12.2 $ & $7.4 \pm 15.0 $ \\
\hline
HD12561 & 199.10 & -73.04 & B6V& 204 & 0.009 & $28.3 \pm 8.7 $ & \\
\hline
HD20404 & 184.93 & -47.95 & B8 & 296 & 0.017 & $33.6 \pm 12.8 $ & $75.3 \pm 4.0$ \\
\hline
HD24446 & 196.02 & -42.31 & B9 & 602 & 0.024 & $14.0 \pm 6.2 $ & $36.2 \pm 8.0 $ \\
\hline
HD30397 & 235.96 & -39.75 & A0V& 203 & 0.002 & $15.0 \pm 6.4 $ & \\
\hline
HD30963 & 208.57 & -31.24 & B9 & 240 & 0.016 & $20.4 \pm 5.9 $ & \\
\hline
HD32043 & 205.56 & -27.57 & B9 & 338 & 0.057 & $45.9 \pm 8.8 $ & $45.4 \pm 19.8 $ \\
\hline
HD33244 & 286.28 & -33.38 & B9.5V& 273 & 0.050 & $39.8\pm 8.4 $ & $85.0 \pm 10.2 $ \\
\hline
HD37104 & 219.11 & -23.74 & B5IV/V & 292 & 0.026 & $18.2 \pm 5.5 $ & \\
\hline
HD37971 & 220.74 & -22.75 & B4/B5III & 565 & 0.030 & $6.2 \pm 9.3 $ & \\
\hline
HD38602 & 290.68 & -30.22 & B8III& 262 & 0.100 & $52.1\pm 10.0 $ & $185.8 \pm 11.9 $ \\
\hline
HD41814 & 218.04 & -14.90 & B3V& 338 & 0.021 & $15.3 \pm 8.5 $ & \\
\hline
HD42849 & 220.94 & -14.70 & B9.5III& 351 & 0.049 & $37.2 \pm 11.1 $ & $54.3 \pm 15.5 $ \\
\hline
HD44533 & 284.47 & -28.52 & B8V& 292 & 0.064 & $22.5 \pm 17.0 $ & $99.8 \pm 24.7 $ \\
\hline
HD44737 & 252.62 & -23.84 & B7V& 746 & 0.010 & $24.3 \pm 6.8 $ & \\
\hline
HD44996 & 221.58 & -11.80 & B4V& 289 & 0.095 & $53.5 \pm 11.5 $ & $130.3 \pm 30.0 $ \\
\hline
HD45040 & 294.17 & -28.30 & B9IV/V & 196 & 0.066 & $31.3\pm 9.1 $ & $96.5 \pm 13.0 $ \\
\hline
HD45098 & 244.76 & -21.11 & B5V& 565 & 0.039 & $31.9 \pm 9.6 $ & \\
\hline
HD46976 & 279.25 & -27.05 & B9V& 364 & 0.014 & & $5.3 \pm 16.3 $ \\
\hline
HD48150 & 252.27 & -20.25 & B3V& 485 & 0.046 & $12.9 \pm 7.8 $ & \\
\hline
HD48872 & 229.32 & -9.99 & B5III/IV & 336 & 0.044 & $22.9 \pm 8.1 $ & $60.3 \pm 13.4 $ \\
\hline
HD49336 & 247.13 & -17.18 & B4Vne& 407 & 0.050 & $10.1 \pm 5.4 $ & \\
\hline
HD49481 & 219.50 & -4.21 & B8 & 365 & 0.033 & $36.0 \pm 6.3 $ & \\
\hline
HD49573 & 224.07 & -6.47 & B8II/III & 395 & 0.060 & $67.4 \pm 5.2 $ & $172.3 \pm 23.5 $ \\
\hline
HD51876 & 228.05 & -5.75 & B9IIw& 333 & 0.064 & $30.0 \pm 17.0 $ & $141.7 \pm 19.7 $ \\
\hline
HD52266 & 219.13 & -0.68 & O9V& 552 & 0.295 & $179.1 \pm 11.5 $ & $518.4 \pm 28.4 $ \\
\hline
HD52849 & 235.09 & -8.32 & B3IV & 1351 & 0.029 & $15.5 \pm 9.1 $ & \\
\hline
HD55523 & 239.64 & -7.95 & B3III& 373 & 0.020 & $18.7 \pm 10.7 $ & \\
\hline
HD57139 & 231.65 & -1.97 & B5II/III & 357 & 0.149 & $99.4 \pm 12.6 $ & $179.7 \pm 8.8 $ \\
\hline
HD60098 & 249.46 & -8.29 & B4V& 248 & 0.043 & $42.5 \pm 5.7 $ & \\
\hline
HD60102 & 296.77 & -26.51 & B9.2/A0V & 207 & 0.085 & $34.1 \pm 13.0 $ & $119.2 \pm 21.6 $ \\
\hline
HD60325 & 230.45 & 2.52 & B2II & 617 & 0.180 & $112.8 \pm 12.4 $ & $254.5 \pm 30.0 $ \\
\hline
HD60929 & 257.11 & -11.45 & A0V& 196 & 0.014 & $22.5 \pm 12.2 $ & \\
\hline
HD61554 & 234.92 & 1.61 & B6V& 257 & 0.072 & $47.4 \pm 5.6 $ & $114.1 \pm 24.0 $ \\
\hline
HD63112 & 230.59 & 6.17 & B9III& 228 & 0.034 & $35.2 \pm 15.5 $ & $106.9 \pm 6.5 $ \\
\hline
HD63868 & 255.17 & -7.41 & B3V& 327 & 0.033 & $19.6 \pm 11.5 $ & \\
\hline
HD65322 & 301.53 & -27.26 & B8IV & 224 & 0.085 & $53.9\pm 6.8 $ & $199.9 \pm 19.4 $ \\
\hline
HD70948 & 260.67 & -3.45 & B5V& 341 & 0.052 & $21.7\pm 9.5 $ & $55.9 \pm 30.0 $ \\
\hline
HD71019 & 260.37 & -3.14 & B3II/III & & 0.095 & $75.1\pm 8.2 $ & $200.9 \pm 30.0 $\\
\hline
HD71123 & 260.22 & -2.92 & B9III& 415 & 0.062 & $44.0\pm 8.3 $ & $47.0 \pm 18.7 $ \\
\hline
HD71336 & 261.01 & -3.21 & B3III/IV & & 0.055 & $23.8\pm 12.8 $ &  \\
 \hline
HD71518 & 237.77 & 13.43 & B2V& 457 & 0.015 & $23.1 \pm 5.9 $ & \\
\hline
\end{tabular}
\end{center}
\begin{footnotesize}
(*) The error on E(B-V) is assumed to be 0.03 mag
\end{footnotesize}
\label{default}
\end{table*}
}

\onltab{2}{
\addtocounter{table}{-1}
\newpage
\begin{table*}[htdp]
\caption{Stellar data and measurements (cont'd)}
\begin{center}
\begin{tabular}{|l|c|c|c|c|c|c|c|c|c|}
\hline
Star & l ($^{\circ}$) & b ($^{\circ}$) & spectral type & d (pc) & E(B-V)* & EW(5780A) (m\AA\ )  & EW(6284A) (m\AA\ ) \\
\hline
HD73687 & 237.52 & 17.03 & A0V& 233 & 0.050 & $34.9 \pm 10.2 $ & $63.5 \pm 10.2 $ \\
\hline
HD75112 & 256.54 & 5.41 & B4V& 385 & 0.023 & $25.4 \pm 16.6 $ & $72.8 \pm 18.5 $ \\
\hline
HD77640 & 234.39 & 26.27 & A0 & 244 & 0.005 & $14.3 \pm 4.5 $ & \\
\hline
HD77665 & 251.63 & 13.90 & B8V& 365 & 0.085 & $76.6 \pm 7.9 $ & $196.8 \pm 18.0 $ \\
\hline
HD79290 & 255.69 & 13.21 & A0V& 244 & 0.068 & $25.8 \pm 15.8 $ & $137.8 \pm 6.6 $ \\
\hline
HD79420 & 276.74 & -6.40 & B4III& 935 & 0.066 & $49.2 \pm 8.5 $ & $156.7 \pm 15.1 $ \\
\hline
HD82984 & 273.03 & 2.04 & B4IV & 254 & 0.043 & & $69.1 \pm 30.0 $ \\
\hline
HD83153 & 274.21 & 1.01 & B3/B4III & 481 & 0.209 & $82.5\pm 6.5 $ & $249.8 \pm 30.0 $ \\
\hline
HD84201 & 265.12 & 13.07 & B9IV & 249 & 0.049 & $54.5\pm 9.1 $ & $148.1 \pm 9.3 $ \\
\hline
HD85355 & 272.95 & 6.31 & B7III& 258 & 0.028 & & $42.5 \pm 30.0 $ \\
\hline
HD86193 & 264.30 & 18.34 & A1III/IV & 216 & 0.059 & $8.2\pm 9.3 $ & $27.4 \pm 17.2 $ \\
\hline
HD86353 & 278.69 & 0.80 & B7V& 243 & 0.042 & $17.6\pm 11.8 $ & $33.8 \pm 19.6 $ \\
\hline
HD86612 & 259.77 & 24.14 & B5V& 244 & 0.080 & $9.4 \pm 7.7 $ & $58.8 \pm 30.0 $ \\
\hline
HD88025 & 255.27 & 31.80 & A0V& 174 & 0.004 & $9.7 \pm 8.9 $ & $38.9 \pm 10.8 $ \\
\hline
HD92946 & 273.66 & 23.36 & B9.5V& 251 & 0.028 & $15.1\pm 12.8 $ & $37.8 \pm 9.5 $ \\
\hline
HD93331 & 262.21 & 39.33 & B9.5V& 177 & 0.018 & $7.9\pm 7.6 $ & $23.2 \pm 10.2 $ \\
\hline
HD93526 & 263.88 & 38.05 & A0III& 407 & 0.055 & $19.0 \pm 12.7 $ & $73.9 \pm 8.3 $ \\
\hline
HD96124 & 300.19 & -22.33 & A1V& 105 & 0.070 & $7.6\pm 15.1 $ & \\
\hline
HD96838 & 272.11 & 37.25 & A0/A1V & 267 & 0.019 & $11.8\pm 13.7 $ & $62.0 \pm 12.9 $ \\
\hline
HD98867 & 284.62 & 20.58 & B9.5V& 205 & 0.038 & $13.0\pm 6.4 $ & $112.1 \pm 12.3$ \\
\hline
HD103077 & 293.11 & 12.33 & B5V& 366 & 0.046 & $34.8\pm 9.6 $ & $85.7 \pm 19.9 $\\
\hline
HD105078 & 292.64 & 26.27 & B8V& 233 & 0.034 & $6.8 \pm 7.9 $ & \\
\hline
HD105313 & 294.94 & 16.61 & B9V& 292 & 0.057 & $41.2\pm 13.3 $ & $98.1 \pm 12.4 $ \\
\hline
HD105610 & 296.43 & 10.20 & B8II & 565 & 0.062 & & $58.3 \pm 14.7 $ \\
\hline
HD106337 & 294.62 & 25.71 & B6V& 500 & 0.033 & & $12.0 \pm 15.5 $ \\
\hline
HD106461 & 302.68 & -25.56 & B9V& 188 & -0.003 & $15.6 \pm 9.1 $ & \\
\hline
HD107931 & 298.19 & 15.25 & B9V& 163 & 0.031 & $11.4 \pm 6.1$ & \\
\hline
HD108344 & 302.28 & -20.96 & B8V& 220 & 0.132 & $60.3\pm 16.4 $ & $192.7 \pm 30.0 $ \\
\hline
HD108610 & 300.28 & 0.88 & B3IV/V & 424 & 0.130 & $50.6\pm 6.2 $ & $343.8 \pm 17.0 $ \\
\hline
HD108792 & 301.61 & -12.58 & B9V& 272 & 0.210 & $102.2 \pm 17.0 $ & $301.2 \pm 13.0 $ \\
\hline
HD111226 & 301.91 & 38.01 & B8V& 227 & 0.077 & $31.7\pm 7.6 $ & $154.0 \pm 17.5 $ \\
\hline
HD111774 & 303.04 & 23.19 & B8V& 143 & 0.018 & $11.3\pm 6.5 $ & \\
\hline
HD112504 & 305.40 & 53.93 & B9 & 195 & 0.033 & & $27.4 \pm 15.3 $ \\
\hline
HD113709 & 304.09 & -10.75 & B9V& 301 & 0.094 & $55.4\pm 14.4 $ & $209.5 \pm 20.1 $ \\
\hline
HD114243 & 308.69 & 42.55 & A0V& 191 & 0.048 & $30.2\pm 9.0 $ & $133.0 \pm 7.2 $\\
\hline
HD114887 & 304.94 & -7.70 & B4III& 275 & 0.186 & $85.6\pm 10.7 $ & $402.0 \pm 19.8 $ \\
\hline
HD115067 & 308.15 & 25.21 & B8V& 592 & 0.018 & & $5.8 \pm 16.3 $ \\
\hline
HD115088 & 304.18 & -17.17 & B9.5/A0V & 147 & -0.011 & $17.6 \pm 9.3 $ & \\
\hline
HD116226 & 308.31 & 13.98 & B6IV & 649 & 0.071 & $21.7\pm 13.6 $ & $124.7 \pm 22.2 $ \\
\hline
HD119283 & 309.61 & 2.95 & B8V& 256 & 0.087 & $84.1 \pm 12.0 $ & $135.1 \pm 14.7 $ \\
\hline
HD120958 & 315.89 & 22.25 & B3Vne& 1136 & 0.107 & $9.2\pm 12.2 $ & $65.5 \pm 30.0 $ \\
\hline
HD121611 & 321.98 & 37.33 & B9.5V& 204 & 0.018 & & $30.1 \pm 10.8 $ \\
\hline
HD123307 & 328.14 & 42.94 & B9IV/V & 212 & 0.064 & & $164.3 \pm 11.5 $ \\
\hline
HD124182 & 311.21 & -4.63 & B3II/III & 420 & 0.165 & $106.0\pm 17.0 $ & $252.7 \pm 30.0 $ \\
\hline
HD124834 & 309.04 & -12.16 & B3III/IV & 346 & 0.154 & $66.0\pm 13.6 $ & $233.0 \pm 30.0 $ \\
\hline
\end{tabular}
\end{center}
\label{default}
\end{table*}
}

\onltab{3}{
\addtocounter{table}{-1}
\newpage
\begin{table*}[htdp]
\caption{Stellar data and measurements (cont'd)}
\begin{center}
\begin{tabular}{|l|c|c|c|c|c|c|c|c|c|}
\hline
Star & l ($^{\circ}$) & b ($^{\circ}$) & spectral type & d (pc) & E(B-V) * & EW(5780A) (m\AA\ )  & EW(6284A) (m\AA\ ) \\
\hline
HD125007 & 319.45 & 17.49 & B9V& 292 & 0.038 & $16.8\pm 6.6 $ & $29.2 \pm 17.5 $ \\
\hline
HD130158 & 333.34 & 30.31 & B9IV/V & 228 & 0.047 & $47.8\pm 15.2 $ & \\
\hline
HD131058 & 315.03 & -6.06 & B3Vn & 391 & 0.112 & $96.6 \pm 9.5 $ & $327.3 \pm 23.4 $ \\
\hline
HD131919 & 333.44 & 26.16 & B9V& 152 & 0.061 & $33.7 \pm 7.5 $ & $143.2 \pm 15.1 $ \\
\hline
HD132101 & 322.27 & 6.43 & B5V& 293 & 0.109 & $57.3 \pm 9.3 $ & $139.7 \pm 28.0 $ \\
\hline
HD133529 & 337.27 & 27.98 & B7V& 165 & 0.123 & $53.1\pm 10.9 $ & $143.5 \pm 21.8 $ \\
\hline
HD135230 & 344.68 & 33.26 & B9III& 245 & 0.095 & $28.2\pm 9.8 $ & \\
\hline
HD135961 & 311.60 & -16.31 & B9V& 262 & 0.063 & $69.8 \pm 8.5 $ & $136.8 \pm 11.2 $ \\
\hline
HD137366 & 314.86 & -12.58 & B3V& 395 & 0.045 & $43.0 \pm 13.9 $ & \\
\hline
HD139094 & 343.04 & 23.20 & B7V& 316 & 0.204 & & $182.4 \pm 16.7  $ \\
\hline
HD139909 & 353.36 & 31.81 & B9.5V& 179 & 0.121 & $79.1 \pm 9.8 $ & $225.2 \pm 15.1 $ \\
\hline
HD140037 & 340.15 & 18.04 & B5III& 344 & 0.077 & $37.8\pm 10.1 $ & $129.1 \pm 12.2 $ \\
\hline
HD140619 & 330.19 & 4.64 & B9III& 420 & 0.106 & $47.0\pm 7.5 $ & $243.0 \pm 17.9 $ \\
\hline
HD141327 & 340.92 & 16.61 & B9V& 195 & 0.063 & $27.6\pm 12.1 $ & $80.9 \pm 13.7 $ \\
\hline
HD142315 & 348.98 & 23.30 & B9V& 148 & 0.129 & $53.3 \pm 8.2 $ & $251.3 \pm 15.8 $ \\
\hline
HD142805 & 350.41 & 23.81 & A0IV & 213 & 0.216 & $112.3\pm 13.0 $ & $435.4 \pm 14.6 $ \\
\hline
HD143321 & 330.58 & 1.22 & B5V& 198 & 0.175 & $86.5 \pm 14.8 $ & $226.9 \pm 20.5 $ \\
\hline
HD143326 & 313.04 & -18.47 & B8V& 301 & 0.056 & $32.6\pm 8.6 $ & $151.9 \pm 15.3 $ \\
\hline
HD146029 & 352.78 & 20.23 & B9V& 214 & 0.133 & $113.1 \pm 17.0 $ & $244.6 \pm 11.2 $ \\
\hline
HD146295 & 320.21 & -13.01 & B8/B9V & 184 & 0.047 & $30.0 \pm 5.5 $ & $93.7 \pm 20.0 $ \\
\hline
HD147932 & 353.72 & 17.71 & B5V& 135 & 0.495 & $197.6 \pm 10.4 $ & $330.5 \pm 27.1 $ \\
\hline
HD149425 & 342.45 & 4.71 & B9V& 181 & 0.191 & $94.4\pm 16.2 $ & $192.2 \pm 22.6 $ \\
\hline
HD150548 & 328.09 & -9.71 & B3V& 259 & 0.087 & $47.1\pm 9.8 $ & $111.8 \pm 19.5 $ \\
\hline
HD151884 & 3.23 & 17.36 & B5V& 243 & 0.438 & $213.6\pm 13.4 $ & $518.4 \pm 30.0 $ \\
\hline
HD152565 & 310.51 & -23.63 & B6IV & 265 & 0.076 & $50.6\pm 7.4 $ & $131.9 \pm 26.8 $ \\
\hline
HD156905 & 332.55 & -12.51 & B4III& 621 & 0.062 & $47.7\pm 6.2 $ & $135.4 \pm 18.1 $ \\
\hline
HD157524 & 328.99 & -15.24 & B7.5V& 256 & 0.046 & $27.9 \pm 12.8 $ & $88.7 \pm 24.3 $ \\
\hline
HD163071 & 336.41 & -15.61 & B4III& 901 & 0.096 & $91.0\pm 8.5 $ & $249.7 \pm 22.5 $ \\
\hline
HD164776 & 351.76 & -9.21 & B5Vn...& 238 & 0.059 & $32.7 \pm 16.5 $ & $52.8 \pm 30.0 $ \\
\hline
HD165052 & 6.12 & -1.48 & O6.5V((f))(n) & 6667 & 0.425 & $183.6 \pm 8.3 $ & $506.3 \pm 30.0 $ \\
\hline
HD165365 & 2.83 & -3.72 & B7/B8III & 366 & 0.092 & $57.7\pm 6.7 $ & $232.4 \pm 15.7 $ \\
\hline
HD165861 & 323.56 & -22.54 & B7.5II-III& 485 & 0.091 & $63.8\pm 10.8 $ & $206.8 \pm 23.7 $ \\
\hline
HD167806 & 334.82 & -19.54 & B2V& 314 & 0.085 & $54.4 \pm 17.0 $ & $135.0 \pm 30.0 $ \\
\hline
HD171577 & 352.32 & -15.76 & B9V& 172 & 0.054 & & $116.1 \pm 18.1 $ \\
\hline
HD171722 & 324.69 & -24.68 & B9V& 258 & 0.034 & $21.8 \pm 8.6 $ & $83.3 \pm 23.7 $ \\
\hline
HD171957 & 18.95 & -3.44 & B8II/III & 247 & 0.327 & $165.8\pm 14.0 $ & $463.6 \pm 18.0 $ \\
\hline
HD172016 & 0.76 & -12.54 & B9.5V& 207 & 0.059 & $29.1 \pm 9.7 $ & $90.3 \pm 13.9 $ \\
\hline
HD172882 & 313.34 & -26.93 & A0V& 309 & 0.101 & $57.2\pm 14.1 $ & $199.8 \pm 17.5 $\\
\hline
HD173545 & 321.22 & -26.07 & B9Vn...& 247 & 0.001 & & $64.9 \pm 27.9 $ \\
\hline
HD176853 & 24.59 & -7.31 & B2V& 216 & 0.438 & $185.8\pm 17.0 $ & $471.8 \pm 30.0 $ \\
\hline
HD177481 & 335.89 & -25.46 & B9V& 439 & 0.076 & $52.4\pm 17.0 $ & $100.7 \pm 20.5 $ \\
\hline
HD179029 & 0.14 & -20.04 & B5V& 188 & 0.262 & $55.1 \pm 8.1 $ & $106.3 \pm 30.0$ \\
\hline
HD182254 & 320.99 & -28.72 & B8/B9Vn& 281 & 0.037 & $10.2 \pm 8.4 $ & \\
\hline
HD185487 & 24.46 & -17.43 & B6III/IV & 205 & 0.086 & $50.0\pm 8.7 $ & $251.6 \pm 14.0 $ \\
\hline
HD186837 & 335.85 & -30.57 & B5V& 285 & 0.007 & $3.9 \pm 8.8 $ & \\
\hline
HD188246 & 355.59 & -29.67 & B8/B9V & 488 & 0.009 & $2.5\pm 10.1 $ & $7.9 \pm 19.3 $ \\
\hline
HD190979 & 23.72 & -25.09 & B8V& 465 & 0.115 & $77.0 \pm 9.5 $ & $278.7 \pm 30.0 $ \\
\hline
HD191091 & 7.65 & -30.05 & B8/B9V & 212 & 0.036 & & $178.1 \pm 14.6 $ \\
\hline
HD196413 & 28.45 & -31.00 & B9V& 213 & 0.039 & $11.2\pm 13.1 $ & $58.9 \pm 12.7 $ \\
\hline
HD198534 & 336.62 & -38.47 & A0IV & 355 & 0.018 & $11.0\pm 12.1 $ & $10.2 \pm 10.0 $ \\
\hline
HD198648 & 27.10 & -35.10 & B9V& 211 & 0.025 & $19.1 \pm 6.7 $ & \\
\hline
HD201317 & 357.66 & -42.94 & B8V& 218 & 0.018 & $6.6\pm 10.2 $ & $5.7 \pm 18.4 $ \\
\hline
HD202025 & 26.08 & -40.74 & A0V& 120 & 0.003 & $5.7\pm 14.5 $ & $12.5 \pm 8.1 $ \\
\hline
HD204220 & 38.21 & -40.65 & B9III/IV & 293 & 0.036 & $33.8\pm 11.6 $ & $19.7 \pm 23.3 $ \\
\hline
HD205348 & 322.56 & -39.32 & B8V& 191 & 0.005 & $4.7 \pm 8.4$ & \\
\hline
HD209386 & 19.37 & -53.18 & B8V& 279 & 0.012 & $1.2\pm 7.4 $ & $0.7 \pm 13.1 $ \\
\hline
HD215047 & 52.30 & -56.52 & B9IV & 207 & 0.042 & & $118.6 \pm 17.5 $ \\
\hline
HD225264 & 16.47 & -79.46 & A0V& 220 & 0.001 & $15.3 \pm 13.3 $ & \\
\hline
\end{tabular}
\end{center}
\label{default}
\end{table*}
}
%\end{appendix}

\end{document}